\documentclass[12pt]{article}
\usepackage{amsmath,amssymb,color}
\usepackage{dsfont}
\usepackage{cite,epsf}
\usepackage{pstricks}
\usepackage{color}

\usepackage{graphicx,array} 
\newcommand{\be}{\begin{equation}}
\newcommand{\ee}{\end{equation}}
\newcommand{\beq}{\begin{equation}}
\newcommand{\eeq}{\end{equation}}
\newcommand{\bea}{\begin{eqnarray}}
\newcommand{\eea}{\end{eqnarray}}

\newcommand{\ba}{\begin{eqnarray}}
\newcommand{\ea}{\end{eqnarray}}

\def\sh{\mbox{sinh}}
\def\ch{\mbox{cosh}}

\begin{document}

\begin{titlepage}
\vspace{10pt}
\hfill
{\large\bf HU-EP-13/32}
\vspace{20mm}
\begin{center}

{\Large\bf  Exceptional conformal anomaly of\\[2mm] 
null polygonal Wilson loops}

\vspace{45pt}

{\large Harald Dorn 
{\footnote{dorn@physik.hu-berlin.de
 }}}
\\[15mm]
{\it\ Institut f\"ur Physik der
Humboldt-Universit\"at zu Berlin,}\\
{\it Newtonstra{\ss}e 15, D-12489 Berlin, Germany}\\[4mm]

\vspace{20pt}

\end{center}
\vspace{10pt}
\vspace{40pt}

\centerline{{\bf{Abstract}}}
\vspace*{4mm}
\noindent
We analyse the breaking of conformal invariance for null polygonal Wilson
loops in ${\cal N}=4$ SYM beyond that induced by the UV divergences due to the cusps.
It only shows up in exceptional configurations, where the polygon intersects
the critical light cone of an inversion or a  special conformal transformation.
In comparison with the related study for the Euclidean version by Drukker and Gross,
we find different leading terms both for weak as well as for strong coupling. Hence the conformal anomaly due to intersections
of a null polygon with a critical light cone defines a new universal function of 
the coupling constant.
\vspace*{5mm}
\noindent
   
\end{titlepage}
\newpage


\section{Introduction}
Wilson loops for closed smooth contours in ${\cal N}=4$ super Yang-Mills gauge 
theory are conformally invariant. They have equal values related to the
contours before and after a conformal map. This statement needs some 
specification, because it holds only if both the original as well as the mapped 
contour are
closed in finite domains of space-time. For the Euclidean version the issue
has been analysed by Drukker and Gross \cite{Drukker}. They started from the 
observation that, although an infinite  straight line and a circle
can be mapped to each other by a suitable conformal transformation, their
values are different. They are equal to one (straight line) or equal to a nontrivial
function of the coupling constant (circle). This fact has been interpreted
as an anomaly with respect to inversions on the unit sphere, emerging as soon as
the original closed contour passes the origin of the coordinate system.
Under such an inversion the origin is mapped to infinity, and consequently
the image contour extends up to infinity and is closed only in the conformal
compactification of $\mathbb{R}^4$.

In this paper we are interested in the analogous problem in Minkowski space.
Now inversions on the unit hyperboloid, as well as special conformal transformations,
map a whole critical light cone to infinity. As soon as the original contour
intersects such a critical light cone, its image consists out of one or several
open pieces forming a closed contour via infinity only. \footnote{By this we mean
closed in the conformal compactification of $\mathbb{R}^{1,3}$.}

Our focus will be on a certain subset of contours, those for
null polygons. They are of particular interest due to their role in the
correspondence to scattering amplitudes\cite{alday-malda,Drummond:2007aua,Brandhuber:2007yx,drummond}. 

Now we are faced with another subtlety. 
Such null polygonal Wilson loops already exhibit breaking of conformal invariance
due the UV divergences introduced by their cusps. We have to make a clear
separation between this generic conformal anomaly and the exceptional one,
showing up only in cases where the null polygon intersects a critical light cone.
The former one is present also for infinitesimal conformal transformations and can be controlled by anomalous conformal Ward identities \cite{Drummond:2007au}.
These Ward identities fix the so-called BDS structure \cite{bds}, which depends
on the {\it not} conformally invariant Mandelstam variables $X_{jk}^2=(X_j-X_k)^2$
and allow the freedom of an additional remainder function, depending
on the conformally invariant cross-ratios formed out of the  $X_{jk}^2$.   
This overall structure is valid for all null polygons closed
in a finite domain. As long as we exclude conformal transformations, whose 
critical light cone cuts the polygon under consideration, we 
have at hand a covariant form of the renormalised Wilson loops. 
The change of their values under a conformal transformation originates
in the change of the Mandelstam variables only. Therefore, the separation
of the exceptional anomaly has to include the definition of a covariant
term utilisable in all cases. The anomaly, we are looking for, is then
the additional change in the value of the Wilson loop, beyond that originating from 
the change of the Mandelstam variables in the covariant term. 

Obviously, our analysis will require the parallel treatment of  
Wilson loops for both null polygons closed in a finite domain as well as those
for null polygons closed only via infinity. Then we have to keep in mind,
that the Mandelstam variables yield an up to isometries unique characterisation
only for the ordered sets of null separated points, which serve as vertices. 
Each pair of consecutive vertices $(X_j,X_{j+1})$ defines a null geodesic.
In forming a null polygon, one then has the option to choose as the related
edge either the finite part connecting $X_j$ and $X_{j+1}$ or the complementary part connecting them via infinity. Therefore, to each set of $n$ vertices and their corresponding
Mandelstam variables belong $2^n$ inequivalent null polygons. An
analysis of the related classification of conformally equivalent null 
polygons has been given in \cite{Dorn:2012cn}.

Section 2 will be devoted to the study of the anomaly in lowest order of field
theoretical perturbation theory. The behaviour at strong coupling via $AdS$/CFT
is studied in section 3. Several technical details are put into two appendices. 
\section{Anomaly at weak coupling}
Analysing the transformation properties of the gluon propagator, Drukker and Gross
pinned down the 1-loop anomaly with respect to an inversion on the 
Euclidean unit sphere to the integral \cite{Drukker}
\beq
-\frac{1}{16\pi ^2}\int dx_1^{\mu}\int dx_2^{\nu}~\partial ^1_{\mu}\left (\mbox{log}
\Big (\frac{(x_1-x_2)^2}{\vert x_1\vert }\Big )~\frac{2x_{2\nu}}{x_2^2}\right )~.
\label{dg}
\eeq
As an integral along a closed contour it is zero formally, but due to the singular
behaviour at $x_1=x_2=0$ it gives a non-vanishing contribution.
Let us consider its Minkowski version for a smooth contour crossing the light cone, centered at the origin, at the point $z\neq 0$, but $z^2=0$. For the moment it should have 
a time-like or space-like tangent vector $t$ at the crossing ($t^2 =\pm 1$). Then with $x_j=z+\sigma _j t+{\cal O}(\sigma_j^2), ~(j=1,2)$ we get for the factor
$$t^{\nu}~\frac{ x_{2\nu}}{x_2^2}~=~\frac{tz+\dots}{2tz\sigma_2+\dots}~=~\frac{1}{2\sigma _2}+{\cal O}(1) ~.$$
This is in contrast to the situation for passing the origin (Euclidean case), where
$z=0$ and therefore
$$t^{\nu}~\frac{ x_{2\nu}}{x_2^2}~=~\frac{t^2\sigma_2+\dots}{t^2\sigma_2^2+\dots}~=~\frac{1}{\sigma _2}+{\cal O}(1) ~.$$
The relative factor $\frac{1}{2}$ in the two above formulas turns out to be the only 
difference in the Minkowskian versus Euclidean analysis. Therefore, we can expect
that the one-loop anomaly, caused by a single crossing of a critical light cone of a 
Minkowskian inversion or a special conformal transformation, is just one half of
the Euclidean anomaly, as long as the tangent vector of the contour at the crossing point is not light-like. 

Since for a closed contour the number of crossings is necessarily even, one could even say that for generic smooth contours  the anomaly situation at one-loop level is in some sense the same as in the Euclidean case.

However it is not clear, whether this statement is still true in the limit
of a light-like tangent vector at the crossing. If beyond that yet a whole 
interval on the contour is light-like around the crossing
point, the singularity structure of \eqref{dg} becomes nonlocal. Furthermore, for
only piecewise smooth contours one has to disentangle the effect from that of
UV divergences due to cusps.  
 
Therefore, we have to perform a careful diagrammatic analysis of Wilson loops
for null polygons closed only via infinity. We will do this for the
case of four vertices with three of them connected by finite edges. 
In fig.\ref{cone-cuts} of the next section one can see an illustration of
how such a situation arises after a conformal map.

\begin{figure}[h!]
 \centering
 \includegraphics[width=10cm]{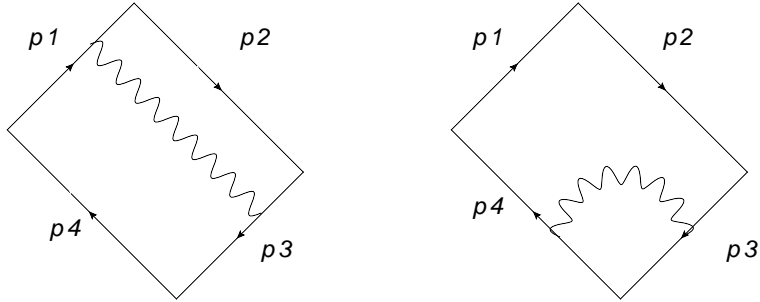} 
\caption{\it Two types of contributions for the finitely closed tetragon, on the left $J_{13}$, on the right $J_{34}$.} 
\label{standard}
\end{figure}

We denote the vertices by $X_1$ to $X_4$, define $p_j:=X_{j+1}-X_j$ and the
Mandelstam variables as usual by
\beq
s~:=~(p_1+p_2)^2~,~~~t~:=~(p_2+p_3)^2~. \label{mandel}
\eeq   
Let us first recall the relevant integrals, see fig.\ref{standard} for a null tetragon closed within a finite domain, see e.g. \cite{Drummond:2007aua,Brandhuber:2007yx}, $~~a:=\frac{g^2N}{8\pi ^2}~\pi^{\epsilon}~\Gamma (1-\epsilon)$, $~~\epsilon$ parameter of dimensional regularisation, $\mu$ RG scale
\bea
J_{13}&=&\frac{a}{4}~\Big (\big (\mbox{log}~\frac{s}{t}\big )^2~+~\pi ^2\Big)~,\nonumber\\
J_{34}&=& -\frac{a}{2}~\frac{1}{\epsilon ^2}~\Big (\frac{-2p_3p_4}{\mu ^2}\Big )^{\epsilon}~.\label{int-standard} 
\eea
These formulas are valid for all signs of the Mandelstam variables and yield 
imaginary parts in case $s$ or $t$ become time-like. Adding the remaining $J_{ik}$
one has altogether
\beq
\mbox{log}~ W(s,t)~=~a~\left (-~\frac{1}{\epsilon ^2}\Big (\Big (\frac{-s}{\mu ^2}\Big )^{\epsilon}+\Big (\frac{-t}{\mu ^2}\Big )^{\epsilon}\Big )~+~
\frac{1}{2}~\Big (\mbox{log}~\frac{s}{t}\Big )^2~+\frac{\pi ^2}{2}\right )~+~{\cal O}(a^2)~.\label{w-standard}
\eeq

\begin{figure}[h!]
 \centering
 \includegraphics[width=10cm]{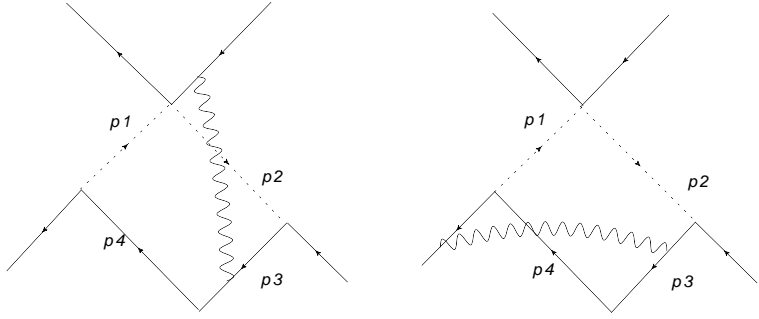} 
\caption{\it The two contributions to the counterpart of the first diagram in fig.\ref{standard}, i.e. integration along the $p_1$-edge replaced by that along its complement via infinity.} 
\label{open-opposite}
\end{figure}

Now we turn to the Wilson loop for a case with $s<0$, $t>0$ and the null polygon going from $X_1$ to $X_2$ via infinity, from $X_2$ to $X_3$ also via infinity but then
with the finite zigzag from $X_3$ via $X_4$ back to $X_1$. The counterpart to
$J_{13}$ is then $\hat J_{13}=\hat J_{13}^{(1)}+\hat J_{13}^{(2)}$, with the two
summands related to the two diagrams of fig.\ref{open-opposite} and given by
\bea
\hat J_{13}^{(1)}&=&\frac{a}{2}~(s+t)~\int _0^{\infty}d\sigma _1\int _0^1d\sigma _2~\big (\sigma _1(1-\sigma _2)~s-\sigma _2(1+\sigma _1)~t-i\varepsilon\big )^{-1}~,\nonumber\\
\hat J_{13}^{(2)}&=&\frac{a}{2}~(s+t)~\int _0^{\infty}d\sigma _1\int _0^1d\sigma _2~\big (-(\sigma_1+1)(1-\sigma _2)~s+\sigma _1\sigma_2 ~t+i\varepsilon\big )^{-1}~.\label{int-open-opposite}
\eea

Let us look in some detail on the crucial differences in comparison to the calculation of $J_{13}$ for $s<0,~t>0$. There one hits the UV light cone singularity, where 
the integrand changes its sign. The $i\varepsilon$-prescription generates an imaginary part and the principal value rule
for the real part. Here one has no UV problem, but instead an IR problem. For this the $i\varepsilon$ is irrelevant. The integrand is negative for the first integral
and positive for the second one. With a symmetric IR cutoff for the $\sigma _1$-integration we get a finite pure real result (for more details see appendix A)  
\beq
\hat J_{13}~=~\frac{a}{4}~\big (\mbox{log}~\left\vert\frac{s}{t}\right\vert\big )^2~.\label{hatJ13}
\eeq
Obviously we also have $\hat J_{24}=\hat J_{13}$.

\begin{figure}[h!]
 \centering
 \includegraphics[width=10cm]{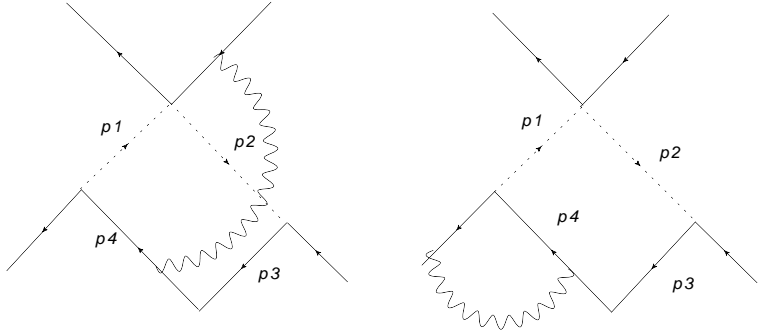} 
\caption{\it The two parts of the cusp contribution for the edge $p_4$ with
the complement to $p_1$ via infinity. } 
\label{open-cusp}
\end{figure}

There are three types of cusp contributions. First of all
\beq
\hat J_{34}~=~ J_{34}.~\label{34}
\eeq  
$\hat J_{14}$ and $\hat J_{23}$ are of the same type. To $\hat J_{14}$ the two
contributing diagrams are depicted in fig.\ref{open-cusp}.
\bea 
\hat J_{14}&=&\frac{a}{2}~\int _0^1d\sigma _2 ~\left (\Big (\frac{-2p_1p_4}{\mu ^2}\Big )^{\epsilon}\int_1^{\infty}d\sigma _1(\sigma_1\sigma_2-i\varepsilon)^{\epsilon -1} \right .\nonumber\\
&&~~~~~~~~~~~~~~~\left . -~\Big (\frac{2p_1p_4}{\mu ^2}\Big )^{\epsilon}~\int _0^{\infty}d\sigma_1~(\sigma _1\sigma _2+i\varepsilon )^{\epsilon -1}\right )~.\label{hat-J14}
\eea
The $\sigma _1$-integration of the second term is both UV and IR divergent. It is a no-scale integral, which by the standard rules of dimensional regularisation has to be set to zero. Then the evaluation of the first term (as a continuation from the IR-side
of $\epsilon$) gives
\beq 
\hat J_{14}~=~ -\frac{a}{2}~\frac{1}{\epsilon ^2}~\Big (\frac{-t}{\mu ^2}\Big )^{\epsilon}~=~\hat J_{23}~.\label{hat-J-14}
\eeq
Note that this equals $J_{14}$ and $J_{23}$.

\begin{figure}[h!]
 \centering
 \includegraphics[width=10cm]{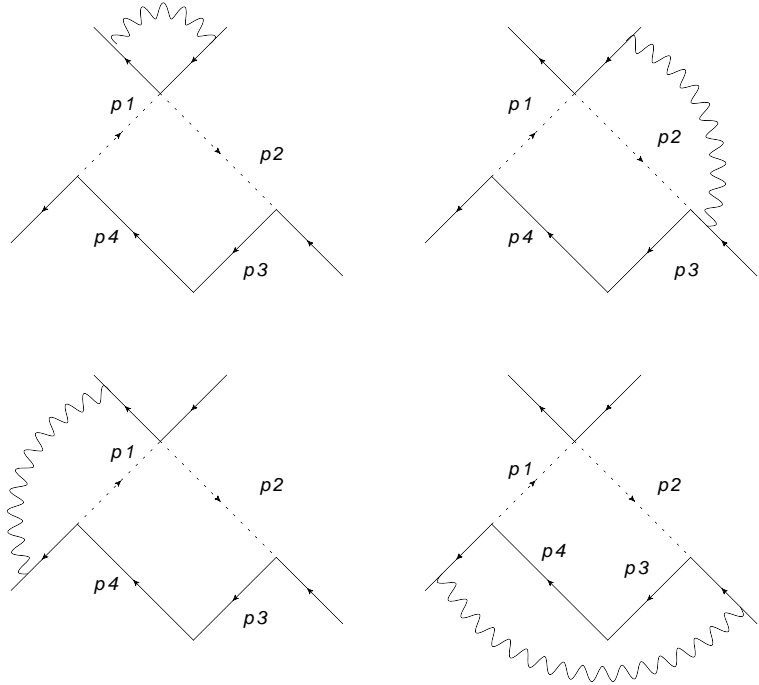} 
\caption{\it The four diagrams contributing to $\hat J_{12}$.} 
\label{open-cusp-2}
\end{figure}

The last type of cusp diagrams, which we have to consider, are those for $\hat J_{12}$.
They are shown in fig.\ref{open-cusp-2}.
The first three diagrams contain no-scale integrals and can be set to zero.
The last diagram yields
\beq
\hat J_{12}~=~ -\frac{a}{2}~\frac{1}{\epsilon ^2}~\Big (\frac{-s}{\mu ^2}\Big )^{\epsilon}~.\label{hat-J-12}
\eeq
Collecting the results from \eqref{hatJ13},\eqref{34},\eqref{hat-J-14} and \eqref{hat-J-12} we get \footnote{Remember that this formula applies to the situation $s<0,~t>0$ and one vertex connected to the others only via infinity.}
\beq
\mbox{log}~\hat W(s,t)~=~a~\left (-~\frac{1}{\epsilon ^2}\Big (\Big (\frac{-s}{\mu ^2}\Big )^{\epsilon}+\Big (\frac{-t}{\mu ^2}\Big )^{\epsilon}\Big )~+~
\frac{1}{2}~\Big (\mbox{log}~\left\vert\frac{s}{t}\right\vert\Big )^2~\right )~+~{\cal O}(a^2)~.\label{hat-W}
\eeq
\\

Now we have all the ingredients to extract the exceptional anomaly. For this purpose
we start with a null tetragon closed in a finite domain with space-like $s,t<0$.
The related Wilson loop is given by eq.\eqref{w-standard}. If we apply a special 
conformal transformation, whose critical light cone does not cut the original 
tetragon, the Mandelstam variables $s',t'$ of the image have the same sign
as those of the original, and the image is again finitely closed. 
The Wilson loop $W'$ for the image is given by the r.h.s. of \eqref{w-standard} with
$s$ and $t$ replaced by $s'$ and $t'$, i.e. $\mbox{log}~W'=\mbox{log}~W(s',t')$. 
Besides this replacement, no further change takes place, i.e. in this case there is 
no exceptional conformal anomaly.

In contrast, we now apply a special conformal transformation for which one of the
vertices, say $X_2$ is within the critical light cone
\footnote{In fig.\ref{cone-cuts} it is $X_1$. We have
a shift in the numbering of vertices between this section and the next one, but
hope that this does not confuse the reader.} 
and the others outside. Then $W'$ is a Wilson loop for a null tetragon closed
only via infinity. One of the Mandelstam variables has changed its sign 
($s'<0,~t'>0$) and we get from \eqref{hat-W}
\beq
\mbox{log}~W'~=~\mbox{log}~\hat W(s',t')~.\label{W-W-prime}
\eeq

As long as both Mandelstam variables are space-like 
$\mbox{log}\frac{s}{t}=\mbox{log}\left\vert\frac{s}{t}\right\vert$. Then we
can declare \eqref{w-standard} with the just mentioned absolute value option
as $the$ covariant term. This gives the exceptional anomaly ('t Hooft coupling $\lambda$)
\beq
{\cal C}_2~=~-\frac{\pi^2}{2}~a~+~{\cal O}(a^2)~=~-\frac{1}{16}~\lambda ~+~{\cal O}(\lambda ^2)~.\label{weak-anom}
\eeq 
The index $2$ stands for two cuts by the critical light cone. ${\cal C}_2$ is half 
of the
anomaly in the Euclidean case \cite{Drukker} and hence also half of that for
generic smooth contours in Minkowski space, as argued at the beginning of this 
section.

Superficially, there seems to be some arbitrariness in fixing the covariant form.
Why, starting from \eqref{w-standard} and a configuration with space-like $s$ 
and $t$, we insist on using the absolute value option for the argument of the 
squared log term, but use no absolute values in the divergent pole term ?
Other choices would generate an imaginary part for the anomaly, which we want 
to avoid. The absence of an imaginary part for strong coupling will become
manifestly in the next section. 
\section{Anomaly at strong coupling}
Within the $AdS$/CFT dictionary, a Wilson loop at strong coupling is
given by 
\beq
\mbox{log}~W~=~-\frac{\sqrt{\lambda}}{2\pi}~{\cal A}~,\label{wilson-malda}
\eeq
with ${\cal A}$ denoting the area of the minimal surface in $AdS_5$
approaching the contour of the Wilson loop at the conformal boundary of $AdS_5$, \cite{Maldacena:1998im}, see also \cite{Drukker,alday-malda} .
Conformal transformations in  $\mathbb{R}^{1,3}$ act as isometries in $AdS_5$.
Therefore, the area of finite surfaces is invariant. The anomaly we are discussing
is due to the fact, that we have to handle infinitely extended surfaces for which
one can define only regularised areas. Following \cite{alday-malda,Alday:2008cg,Zarembo} we use a cut-off at a constant value of the Poincar$\acute{\mbox{e}}$ coordinate
$r=r_c$. The anomaly under a conformal map is then due to the fact, that 
the image of the boundary of the regularised surface is no longer located at a constant 
value of $r$. 

To set up some notation, we recall the relation between the embedding coordinates
of the $AdS_5$ hyperboloid in $\mathbb{R}^{2,4}$
\beq
Y_0^2+Y_{0'}^2-Y_1^2-Y_2^2-Y_3^2-Y_4^2~=~1\label{hyperboloid}
\eeq  
and  Poincar$\acute{\mbox{e}}$ coordinates:
\beq
Y^{\mu}~=~\frac{x^{\mu}}{r},~~\mu =0,1,2,3~~~~~Y^{0'}+Y^4=\frac{1}{r}~,~~~Y^{0'}-Y^4=
\frac{r^2-x^{\mu}x_{\mu}}{r}~.
\eeq
The whole $AdS_5$ hyperboloid is covered by two Poincar$\acute{\mbox{e}}$ patches,
defined by  $r>0$ and $r<0$. Its conformal boundary is conformal to two copies of  $\mathbb{R}^{1,3}$, corresponding to $r\rightarrow +0$ and  $r\rightarrow -0$, respectively.

We consider a minimal surface in $AdS_5$ with null tetragonal boundary and
parameterised in Poincar$\acute{\mbox{e}}$ coordinates by ($-\infty<\xi ,\eta<\infty$)
\bea
r~=~\frac{\sqrt{2}~a}{\ch \eta+b~\ch\xi}~,&&x^0~=~\frac{a~\sqrt{1+b^2}~\ch\xi}{\ch \eta+b~\ch\xi}~,\nonumber\\
x^1~=~\frac{a~\sh\xi}{\ch \eta+b~\ch\xi}~,&&x^2~=~\frac{a~\sh\eta}{\ch \eta+b~\ch\xi}~,~~~~~x^3~=~0~.\label{surface}
\eea
Here $a>0,~b\geq 0$ are parameters of isometry transformations in $AdS_5$, applied
to the initial configuration ($a=1,~b=0$) of ref. \cite{alday-malda}, given in embedding coordinates as the intersection of $Y_0^2-Y_{0'}^2=Y_1^2-Y_2^2~,~~~ Y_3=Y_4=0$ and the $AdS_5$ hyperboloid.
\footnote{For details see also the $\theta =\frac{\pi}{4}$ case in ref. \cite{Dorn:2009hs}.}
This surface is completely located in one Poincar$\acute{\mbox{e}}$ patch ($r>0$).

The vertices of the null tetragon approached for $r\rightarrow +0$ are obtained 
from \eqref{surface} as follows: $X_1$ and $X_3$ in the limits $\xi\rightarrow\pm\infty, ~\eta$ fix and  $X_2$ and $X_4$
in the limits $\eta\rightarrow\pm\infty,~\xi$ fix  
\footnote{In the notation below we neglect the coordinate $x_3$,
which is always equal to zero.}
\bea
X_{1}~=~\Big (\frac{a~\sqrt{1+b^2}}{b},~\frac{a}{b},~0 \Big  )&,&~~~X_{2}~=~(0,~0,~a)~,\nonumber\\
X_{3}~=~\Big (\frac{a~\sqrt{1+b^2}}{b},~-\frac{a}{b},~0\Big  )&,&~~~X_{4}~=~(0,~0,~-a) ~.\label{vertices}
\eea
The related Mandelstam variables are
\beq
s~=~(X_{1}-X_{3})^2~=~-4~\frac{a^2}{b^2}~,~~~~~t~=~(X_{2}-X_{4})^2~=~-4a^2~.
\label{mandelstam}
\eeq

Now we want to act with  a special conformal transformation in Minkowski space
$\mathbb{R}^{1,3}$. Its extension to an isometry of $AdS_5$ is given by
\beq
x'^{\mu}~=~\frac{x^{\mu}+c^{\mu}(x^2-r^2)}{1+2cx+c^2(x^2-r^2)}~,~~~r'~=~\frac{r}{1+2cx+c^2(x^2-r^2)}~.\label{conf-trafo}
\eeq
In $\mathbb{R}^{1,3}$ it maps a whole light cone, whose tip is located at $-c^{\mu}/c^2,$ to infinity. If this critical light cone cuts an edge of the tetragon, the image of this edge is not completely in a finite domain of Minkowski space, but it is just the piece of the null geodesic connecting the images of the two adjacent vertices
via infinity.

\begin{figure}[h!]
 \centering
 \includegraphics[width=12cm]{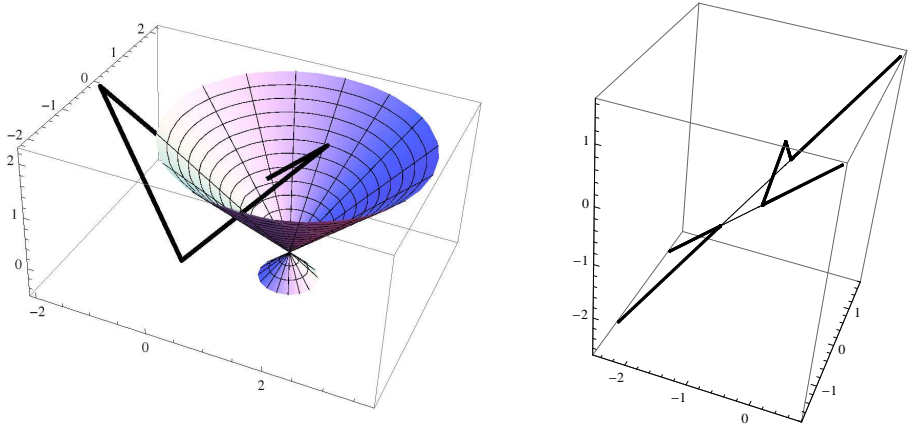} 
\caption{\it Tetragon with a=1, b=0.5 before and after a special conformal transformation. The critical light cone has h=1.5.} 
\label{cone-cuts}
\end{figure}

Let us choose 
\beq
c~=~\big (0,\frac{1}{h},0\big )\label{c}~,
\eeq
then the tip of the critical light cone is located at $x^1=h$ on the $x^1$-axis. 
Varying $h$ we can study three cases:\footnote{Our set-up is symmetric under
$h\rightarrow -h$. In the following for convenience we always assume $h>0$.\label{f1}}\\[2mm]
\underline{\it Case A}: no edge cut,
\beq
h^2~>~\frac{a^2}{b^2}~(\sqrt{1+b^2}+1)^2~.\label{A}
\eeq
\underline{\it Case B}: two adjacent edges cut,
\beq
\frac{a^2}{b^2}~(\sqrt{1+b^2}+1)^2~>~h^2~>~\frac{a^2}{b^2}~(\sqrt{1+b^2}-1)^2~.\label{B} 
\eeq
\underline{\it Case C}: all edges cut,
\beq
\frac{a^2}{b^2}~(\sqrt{1+b^2}-1)^2~>~h^2~.\label{C} 
\eeq
The situation in {\it case B} and the image of the tetragon under the special
conformal map is illustrated in fig.\ref{cone-cuts}.

The images of the vertices \eqref{vertices} are
\bea
X'_1=\frac{h^2}{bh^2-2ah-a^2b}\Big (a\sqrt{1+b^2},~\frac{ah+a^2b}{h},~0\Big),~~X'_2=\frac{h^2}{h^2+a^2}\big (0,~-\frac{a^2}{h},~a\big ),~~~\\
X'_3=\frac{h^2}{bh^2+2ah-a^2b}\Big (a\sqrt{1+b^2},~\frac{a^2b-ah}{h},~0\Big),~~X'_4=\frac{h^2}{h^2+a^2}\big (0,~-\frac{a^2}{h},~-a\big ).\nonumber\label{mapped-vertices} 
\eea
The Mandelstam variables for the image are
\beq
s'~=~-4~\frac{a^2}{b^2}~\frac{b^2h^4}{b^2(h^2-a^2)^2-4a^2h^2}~,~~~t'~=~-4a^2~\frac{h^4}{(h^2+a^2)^2}~.\label{mapped-Mandelstam}
\eeq
While $t'$, like the original $t$, is space-like for all three {\it cases A,B,C},
we observe
\beq
s'~<0~~~\mbox{in {\it case A} and } \mbox{\it case C}~,~~~~~~s'>0~~~\mbox{in {\it case B}}~.
\label{signs-mapped-s}
\eeq

\begin{figure}[h!]
 \centering
 \includegraphics[width=15cm]{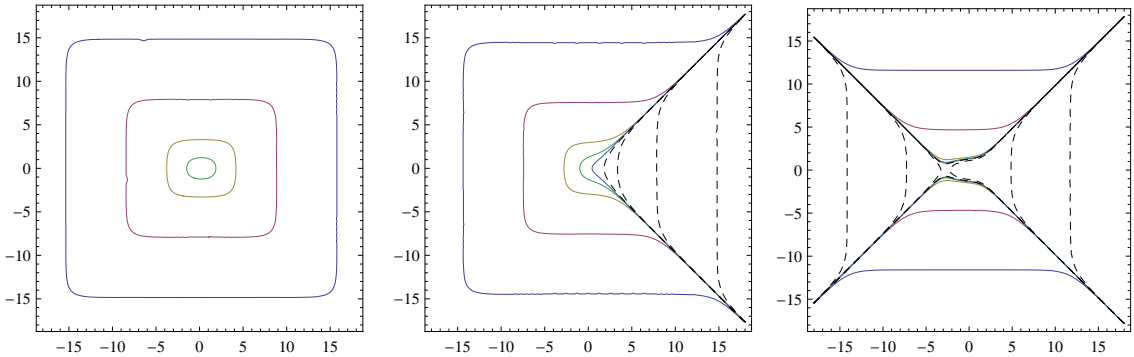} 
\caption{\it Contour lines of $r'$ in the $(\xi ,\eta)$-plane for $a=1,~b=0.5$ and
$h=20$ (case A), $h=1.5$ (case B) and $h=0.2$ (case C). Solid lines are for positive $r'$, dashed lines for negative $r'$. The chosen values for $\vert r'\vert$ are: $10^{-6},10^{-3}, 0.1,0.6$ and in cases B and C also $\infty$.  } 
\label{contours}
\end{figure}

The $r$-coordinate of the image of the surface \eqref{surface} under \eqref{conf-trafo},\eqref{c} is
\beq
r'~=~\frac{\sqrt{2}}{{\cal F}(\xi ,\eta)}~,\label{r-prime-F}
\eeq
with
\beq
{\cal F}(\xi ,\eta):=~\frac{h^2+a^2}{ah^2}~\ch ~\eta~+~b~\frac{h^2-a^2}{ah^2}\ch ~\xi~-~\frac{2}{h}~\sh ~\xi~.\label{F}
\eeq
By inspection,  in {\it case A} we find $r'$  to be positive in the whole  $(\xi ,\eta)$-plane. For {\it cases B} and {C} it takes both signs. Furthermore, it goes to
$\pm\infty$ along one and two lines in {\it case B} and {\it case C}, respectively. Contour lines for three explicit examples are shown in fig.\ref{contours}.

This pattern reflects the fact that in {\it case A} the whole image
of the original surface is still in the same  Poincar$\acute{\mbox{e}}$ patch, while in {\it cases B} and {\it C} it has parts in both patches. At the lines $r'\rightarrow\pm\infty$ it just changes the patch. Of course the mapped surface, as an embedding in $AdS_5$, is smoothly on this lines, since it is obtained by an isometry transformation of the original one. In {\it cases B} and {\it C} its boundary is still a closed null tetragon in the union of the two conformal  $\mathbb{R}^{1,3}$ copies constituting the boundary of $AdS_5$. The identification of antipodes $Y\sim -Y$ on the hyperboloid \eqref{hyperboloid} induces the standard conformal compactification 
of  $\mathbb{R}^{1,3}$, then the boundary tetragon closes only via infinity.

\begin{figure}[h!]
 \centering
 \includegraphics[width=15cm]{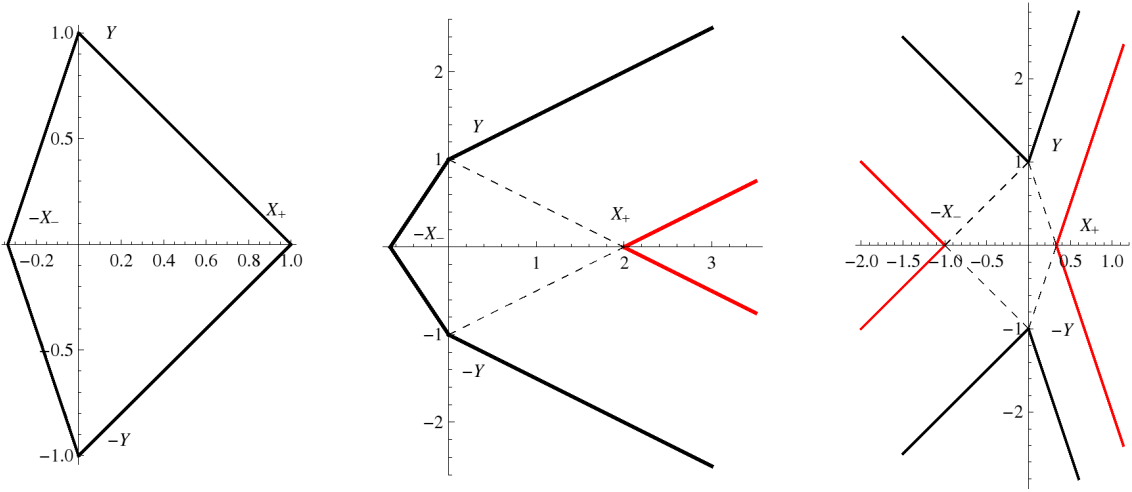} 
\caption{\it Examples for the asymptotic integration region $B_0$ for cases A,B,C with $\beta =1$ and $\alpha = 2,~\frac{1}{2},~-2$, respectively. The black and red lines correspond to  Poincar$\acute{\mbox{e}}$ patches $r>0$ and $r<0$, respectively. The dashed lines indicate the subdivision into a quadrilateral and the strips. The axes denote the coordinates $x$ and $y$. } 
\label{asintreg}
\end{figure}

To define the regularised area of the mapped surface ${\cal A}_{r_c}$, we cut off 
its parts $0<r'<r_c$ in the original patch and $-r_c <r'<0$ in the other one.   The induced metric on the space-like minimal surface \eqref{surface} is
\beq
d \mbox{s}^2~=~-\frac{1}{2}~(d\xi ^2~+~d\eta ^2)~.\label{ind-metric}
\eeq
Using $\xi$ and $\eta$ also as coordinates on the image of this surface
under the $AdS_5$ isometry  \eqref{conf-trafo},\eqref{c} we get
\beq
{\cal A}_{r_c}~=~\frac{1}{2}~\int _{{\cal B} _{r_c}}~d\xi~d\eta ~,~~~~~~~
 {\cal B}_{r_c}~=\Big \{~(\xi,\eta)~\Big \vert  ~\vert {\cal F}(\xi ,\eta)\vert ~<\frac{\sqrt{2}}{r_c}~\Big \}~.
\label{reg-area}
\eeq
Note that this formula holds in all three {\it cases A,B,C}. 
With the abbreviations
\beq
\alpha :=~b~\frac{h^2-a^2}{2ah}~,~~~\beta :=~\frac{h^2+a^2}{2ah}~,~~~\delta :=~r_c ~\frac{\sqrt{2}}{h}~,\label{abcd}
\eeq
and the change of coordinates  $x=\frac{r_c ~\sqrt{2}}{h}~\sh ~\xi ,~y=\frac{r_c ~\sqrt{2}}{h}  ~\sh ~\eta $, we arrive at
\beq
{\cal A}_{r_c}~=~\frac{1}{2}~\int _{B _{r_c}}~\frac{dx~dy}{\sqrt{\delta ^2+x^2}~
\sqrt{\delta ^2+y^2}}~+~o(1) ~,~~~~~~~
B_{r_c}~=\Big \{~(x,y)~\Big \vert ~\vert  F_{r_c}(x,y)\vert ~<1~\Big \}~,\label{A-hatF}
\eeq
where
\beq
F_{r_c}(x,y):=~\alpha ~\sqrt{\delta ^2+x^2}~+~\beta ~\sqrt{\delta ^2+y^2}~- ~x~.\label{F-approx}
\eeq
The coefficient $\beta$ is always positive (see \eqref{abcd} and footnote \ref{f1}) and equal to
\beq
\beta~=~\sqrt{-\frac{h^2}{t'}}~.\label{bMandel}
\eeq
For $\alpha$
one finds from \eqref{A}-\eqref{C} and \eqref{mapped-Mandelstam}
\bea
\underline{\it Case~ A}:&~&\alpha ~=~\sqrt{ 1-\frac{h^2}{s'}}~>1~,\nonumber\\
\underline{\it Case~ B}:&~&-1~<~\alpha ~=~\mbox{sgn}(h-a)~\sqrt{\Big\vert 1-\frac{h^2}{s'}\Big\vert}~ <~1~,~~~~~~~~~~\nonumber\\
\underline{\it Case~ C}:&~&\alpha ~=~-~\sqrt{ 1-\frac{h^2}{s'}} ~<~-1~.\label{signs-a}
\eea
With this information we can characterise the boundary of the asymptotic integration region 
$B_0~=~\lim _{r_c\rightarrow 0}~B_{r_c}$ via 
\beq 
F_0~=~\alpha ~\vert x\vert~+~\beta ~\vert y\vert ~-~x~=~\pm ~1~.\label{asB}
\eeq
This region is shown for all three {\it cases} in fig.\ref{asintreg}.
The marked points in the pictures of fig.\ref{asintreg} are defined by
\beq
X_{\pm}:=~(x_{\pm},0)~=~\Big (\frac{1}{\vert \alpha ~\mp~1\vert}~,~0\Big )~,~~~~~~~~Y:=~(0,y_0)~=~\Big (0,~\frac{1}{\beta}\Big )~.\label{XY}
\eeq
$B_0$ contains always the quadrilateral $(X_+,Y,X_-,-Y)$. In {\it cases B} and {\it C} there is an additional contribution by  two and four semi-infinite strips between
parallel straight lines, respectively. These strips are the asymptotic images of the spikes \footnote{Note that these spikes are infinitely long already for $r_c >0$.} in the
$(\xi,\eta )$-plane (see fig.\ref{contours}) under the map $(\xi,\eta )\rightarrow (x,y)$.
 
Once again we are not interested in contributions to ${\cal A}_{r_c}$ vanishing for
$r_c\rightarrow 0$. Therefore, as argued in appendix B, we can 
in \eqref{A-hatF} replace the integration region $B_{r_c}$ by $B_0$. Then
we get
\bea
\mbox{\underline{\it Case A:}} &~~&{\cal A}_{r_c}~=~I^{\mbox{~\scriptsize quad}}_{r_c}~+~o(1)~,
\nonumber\\
\mbox{\underline{\it Case B:}}  &~~&{\cal A}_{r_c}~=~I^{\mbox{~\scriptsize quad}}_{r_c}~+~2~I^{\mbox{\scriptsize strip}}~+~o(1)~,\nonumber\\
\mbox{\underline{\it Case C:}} &~~&{\cal A}_{r_c}~=~I^{\mbox{~\scriptsize quad}}_{r_c}~+~4~I^{\mbox{\scriptsize strip}}~+~o(1)~,\label{A-quad-strip}
\eea
with $I^{~\mbox{\scriptsize quad}}_{r_c}$ and $I_{\mbox{\scriptsize strip}}$ calculated in appendix B as (see \eqref{Iquad},\eqref{Istrip})
\bea
I^{~\mbox{\scriptsize quad}}_{r_c}&=&\frac{1}{2}~\mbox{log}\Big (\frac{4 x_+x_-}{\delta ^2}\Big )~\mbox{log}\Big (\frac{4~y_0^2}{\delta ^2}\Big )~-~\frac{\pi ^2}{3}~+o(1)~,\\
I^{\mbox{\scriptsize strip}}&=&\frac{\pi ^2}{4}~+o(1)~.
\eea
Relating $\delta$ to $r_c$ by \eqref{abcd} and expressing the coordinates $x_{\pm}$ and $y_0$ of the vertices of the quadrilateral
via \eqref{XY},\eqref{signs-a},\eqref{bMandel} in terms of the Mandelstam variables
$s',t'$, we finally arrive at
\beq
I^{~\mbox{\scriptsize quad}}_{r_c}~=~\frac{1}{4}~\Big (\mbox{log}\frac{2\vert s'\vert}{r_c ^2}\Big )^2~+~\frac{1}{4}~\Big (\mbox{log}\frac{2\vert t'\vert}{r_c ^2}\Big )^2~-~\frac{1}{4}~\Big (\mbox{log}\Big\vert\frac{s'}{t'}\Big\vert\Big )^2
~-~\frac{\pi ^2}{3}~+~o(1)~.\label{Ifinal}
\eeq
Summarising, the regularised area is given by
\beq
{\cal A}_{r_c}~=~\frac{1}{4}~\Big (\mbox{log}\frac{2\vert s'\vert}{r_c ^2}\Big )^2~+~\frac{1}{4}~\Big (\mbox{log}\frac{2\vert t'\vert}{r_c ^2}\Big )^2~-~\frac{1}{4}~\Big (\mbox{log}\Big\vert\frac{s'}{t'}\Big\vert\Big )^2~-~\frac{\pi ^2}{3}~+~n\cdot \frac{\pi ^2}{4}~+~o(1)~,\label{A-reg}
\eeq 
with $n=0,~2,~4$ in {\it cases A,B,C}, respectively.\\ 

In {\it case A} our result for ${\cal A}_{r_c}$ agrees
with \cite{Alday:2008cg},\cite{Dorn:2009hs} after $(s',t')\rightarrow (s,t)$.
\footnote{And taking into account, that there the  Mandelstam variables 
have been defined with respect to the correspondence with scattering amplitudes,
i.e. $(s,t)=4\pi ^2(s,t)\vert _{\mbox{\scriptsize there}}$.} This confirms covariance under all finite conformal transformations mapping closed null tetragons to closed null tetragons, i.e. only the Mandelstam variables of the original configuration have
to be replaced by that of the transformed configuration.  

Beyond that we found an anomaly, showing up as soon as the image of the original
tetragon is no longer closed in a finite region of Minkowski space. This anomaly
for the area ${\cal A}$, relevant for Wilson loops at strong 't Hooft coupling,
is given by the pure number $\frac{\pi ^2}{4}$ for each crossing of the original
null tetragon with the critical light cone of a given finite special conformal
transformation. Due to the clear localisation of the anomaly effect
in the strips of fig.\ref{asintreg} or spikes of fig.\ref{contours} it seems to be obvious, that this pattern is valid also
for higher null polygons. With eq.\eqref{wilson-malda} we get for the exceptional
conformal anomaly of the Wilson loop at strong coupling
\beq
{\cal C}_n~=~-\frac{\pi}{8}~n~\sqrt{\lambda}~+~\mbox{non-leading}~.\label{strong-anom}
\eeq
${\cal C}_2$ has a relative factor $\frac{\pi}{4}$ in comparison with its natural
partner in Euclidean space \cite{Drukker,Zarembo}.

There is still an interesting side remark. The covariant term in the area, present 
in all  {\it cases A,B,C} depends only on the absolute values of the Mandelstam variables. Let us for a 
moment follow a false option: start in the region where both $s$ and $t$ are space-like, replace $(\vert s\vert,\vert t\vert )$ by $(-s,-t)$ and declare this as 
the covariant form for the area. Then one would obtain an imaginary part 
if $s$ and $t$ have
different signs. 

However, this simply reflects the fact, that there is no minimal
surface ending on a null tetragon, closed in a finite region of 
$\mathbb{R}^{1,3}$, with two space-like and two time-like
cusps.  
Instead, our covariant form plus anomaly contribution
yields the area of a minimal surface ending on the null tetragon with the same vertices, but closed via
infinity and having space-like cusps only. 
\section{Conclusions}
We have analysed the conformal transformation properties of a Wilson loop 
in ${\cal N}=4$ SYM for a null tetragon with space-like cusps. As long as the
original null tetragon is not cut by the critical light cone of a special
conformal transformation, the change of its value is only due to the change
of the Mandelstam variables. However if edges are cut, there appears
an additional change, which we called exceptional conformal anomaly.  
At weak coupling we found for two cuts eq.\eqref{weak-anom} and for strong
coupling and two or four cuts eq.\eqref{strong-anom}. 

From the details of its genesis in sections 2 and 3 it seems to be quite 
obviously, that the formulas hold also for higher null polygons.
Furthermore, the
exceptional anomaly is proportional to the number of cuts. Since this
number is necessarily even, the basic entity is
\beq
{\cal C}_2~=~\left \{
\begin{array}{ll}
-\frac{1}{16}~\lambda ~+~{\cal O}(\lambda ^2)~,~~~~~~~~~~\lambda\rightarrow 0~,\\
-\frac{\pi}{4}~\sqrt{\lambda}~+~\mbox{non-leading}~,~~\lambda\rightarrow\infty ~.
\end{array}\right .
\eeq
In comparison with the analogous anomaly in the Euclidean case 
\cite{Drukker,Zarembo} we have a relative factor $1/2$ at weak coupling and $\pi/4$
at strong coupling. Therefore, ${\cal C}_2$ defines a new universal function
of the coupling $\lambda $. Of course it is an interesting open question,
whether it appears also in other context.

A nearby issue, which should be answered by a straightforward adaption of our analysis, 
concerns the even more exceptional situations where the crossing takes places at vertices
or where whole edges are located at a critical light cone.

We also argued at the beginning of section 2, that for Wilson loops for
generic smooth contours (not null at the crossings with a critical light cone)
the anomaly at weak coupling in lowest order is $2{\cal C}_2$, in agreement
with the Euclidean case. 
It would be interesting to further clarify, whether this continues in higher orders.
Also for strong coupling one can analyse the situation, at least for straight lines
and space-like circles with their well-known related minimal surfaces. In the 
Euclidean case under an inversion on the unit
sphere a straight line, not passing the origin, is mapped to a circle. 
In $\mathbb{R}^{1,3}$ under inversion on the unit hyperboloid e.g. a straight 
space-like line crossing
the light cone and its axis is mapped to a space-like hyperbola. For the area related to a 
Euclidean circle one has ${\cal A}=L(1/r_c -1/R)$, with $L=2\pi R$ the circumference 
of a circle of radius $R$ \cite{Zarembo}. In the case of a hyperbola with radius $R$
the circumference is infinite and one is faced with an additional long 
distance (infrared) problem. Handling it similar to that for the straight line
one gets  ${\cal A}=L_{\mbox{\tiny IR}}(1/r_c -1/R)$ with a free IR 
regularisation parameter $L_{\mbox{\tiny IR}}$. 

Our discussion necessarily included Wilson loops for null polygons closed only
via infinity. Further study should clarify, whether they are connected via some
analytic continuation to the finitely closed ones with the same vertices. 
This touches also the question concerning a possible role in the duality to 
scattering amplitudes.
\\[20mm]
\noindent
{\bf Acknowledgement}\\[2mm]
I would like to thank B. Hoare, G. Jorjadze, M. Staudacher, H.S. Yang and \\
K. Zarembo for useful discussions.
\\[10mm]

\section*{Appendix A}
Perhaps the most efficient way to calculate $\hat J_{13}$ is to start with
\beq 
J^{\mbox{\scriptsize total}}_{13}~:=~J_{13}~-~\hat J_{13}~=~\frac{a}{2}(s+t)\int_{-\infty}^{\infty}d\sigma _1\int _0^1d\sigma _2\big (\sigma _1(1-\sigma _2)s+\sigma _2(1-\sigma _1)t-i\varepsilon \big )^{-1}~.\label{J-tot}
\eeq 
That $\hat J_{13}$ contributes with a relative minus sign is due to its opposite
orientation along the contour. The $\sigma_1$-integral is of the type
\beq
\int _{-\infty}^{\infty}dx~\frac{1}{A x-B -i\varepsilon}~=~\frac{i\pi}{\vert A \vert}~.\label{master-int}
\eeq
Here the IR problem has been treated with a symmetric cutoff 
at $x=\pm K,~~K\rightarrow\infty$.
Using \eqref{master-int} we get
\beq
J^{\mbox{\scriptsize total}}_{13}~=~\frac{a}{2}(s+t)~i\pi\int _0^1d\sigma _2~\frac{1}{\vert (1-\sigma _2)s-\sigma _2t\vert }~.
\eeq  
We are interested in the kinematical situation $s<0,~t>0$. Then the $\sigma_2$ integrand has no singularity and we get
\beq
J^{\mbox{\scriptsize total}}_{13}~=~\frac{a}{2}~i\pi ~\mbox{log}\Big (\frac {t}{-s}\Big )~=~\frac{a}{2}~i\pi ~\mbox{log}\Big \vert\frac {t}{s}\Big \vert~.
\eeq
Inserting this together with \eqref{int-standard} into \eqref{J-tot} gives
our formula for $\hat J_{13}$ used in the main text, \eqref{hatJ13}.
\section*{Appendix B}
Here we collect the arguments for the replacement of the integration region
$B_{r_c}$ in \eqref{A-hatF} by $B_0$, defined in \eqref{asB}, as well as some comments on the calculation 
of the integrals $I^{\mbox{~\scriptsize quad}}_{r_c}$ and 
$I^{\mbox{\scriptsize strip}}$
needed in \eqref{A-quad-strip}. 

Let us start with the replacement of $B_{r_c}$ by $B_0$. The positive definite integrand for ${\cal A}_{r_c}$ in \eqref{A-hatF} has a uniform $r_c$-independent
bound outside small circles of radius $R$ around the vertices of the quadrilateral $X_+,Y,Y_-,-Y$. Therefore, the contributions to the integrals due to regions of $B_{r_c}$ and $B_0$ outside these circles differ only by terms vanishing for $r_c\rightarrow 0$. Inside the circles the integrand is bounded by $K/r_c $ with
$K$ some positive constant, depending on the parameters $a,b,h$ and $R$ only. The area between the boundaries of $B_{r_c}$ and $B_0$ inside these circles vanishes
of order $r_c ^2$. Hence also from inside the circles the contributions to the difference of ${\cal A}_{r_c}$ based on $B_{r_c}$ and $B_0$ vanish in the 
limit. 

Let us sketch the argument for the stated $r_c ^2$-behaviour of the
difference area in the $(x,y)$-plane, e.g. near the point $Y$. Here we have to compare
the integral $\int (1+x-\alpha\vert x\vert)dx$ for $B_0$ with ($\delta =\sqrt{2}r_c /h$)
$$\int\sqrt{\big (1+x-\alpha \sqrt{x^2+\delta ^2}\big )^2-\beta ^2\delta ^2}~dx$$ 
for $B_{r_c}$. Around $x=0$ the integrand of the last integral can be expanded as
$$ \big (1+x-\alpha\sqrt{x^2+\delta ^2}\big )\cdot \big (1+O(\delta ^2)\big )~.$$
At least now the necessary integrations become trivial.\\

The quadrilateral integral, used in the main text, is defined by  
\bea
I^{~\mbox{\scriptsize quad}}_{r_c}&:=&\frac{1}{2}~\int _{B_0}~\frac{dx~dy}{\sqrt{\delta ^2+x^2}~\sqrt{\delta ^2+y^2}}~\\
&=& \int _{-x_-}^0~\mbox{log}\left (\frac{y_0(x+x_-)+\sqrt{\delta ^2~x_-^2+y_0^2(x+x_-)^2}}{\delta ~x_-}\right )~\frac{dx}{\sqrt{\delta ^2+x^2}}\nonumber\\
&~&+~ \int ^{x_+}_0~\mbox{log}\left (\frac{y_0(x_+-x)+\sqrt{\delta ^2~x_+^2+y_0^2(x_+-x)^2}}{\delta ~x_+}\right )~\frac{dx}{\sqrt{\delta ^2+x^2}}\nonumber ~.
\eea
Performing now the trivial part of the $x$-integration we get
\bea
I^{~\mbox{\scriptsize quad}}_{r_c}&=&2~\mbox{log}^2\delta ~-~\mbox{log}~\delta ~\Big  (
\mbox{log}(4y_0^2x_+x_-)+O(\delta ^2)\Big )\label{a}\\
&~&+~\mbox{log}~y_0~\mbox{log}(4x_+x_-) +O(\delta ^2) +M(x_+,y_0)+M(x_-,y_0)~,\nonumber
\eea
with
\beq 
M(m,n):=~\int _0^1~\mbox{log}\Big (u+\sqrt{u^2+\delta ^2/n^2}\Big )
\frac{du}{\sqrt{(1-u)^2+\delta ^2/m^2}}~.
\eeq
Replacing the lower integration bound by $\sqrt{\delta}$ we make an error
of order $\sqrt{\delta}~\mbox{log}\delta $ vanishing for $\delta\rightarrow 0$.
Then we can expand the square root under the logarithm in powers of $\delta/u$
and get
\bea
M(m,n)&=&\int _0^1\frac{\mbox{log}(2u)~du}{\sqrt{(1-u)^2+\delta ^2/m^2}}~+~o(1)\nonumber\\
&=&\mbox{log}2~\big (\mbox{log}(2m)-\mbox{log}\delta \big )~-~\frac{\pi ^2}{6}~+~o(1)~.
\eea
Using this in \eqref{a} gives
\beq
I^{~\mbox{\scriptsize quad}}_{r_c}~=~\frac{1}{2}~\mbox{log}\Big (\frac{4 x_+x_-}{\delta ^2}\Big )~\mbox{log}\Big (\frac{4~y_0^2}{\delta ^2}\Big )~-~\frac{\pi ^2}{3}~+o(1)~.\label{Iquad}
\eeq
Finally we calculate \footnote{We write the explicit defining formula for the right upper strips
in fig.\ref{asintreg}. It will become obviously, that the other strips give the
same result.} 
\bea
I^{\mbox{\scriptsize strip}}&:=&\frac{1}{2}~\int_0^{x_+}\frac{dx}{\sqrt{\delta ^2+x^2}}\int _{y_0(1-x/x_+)}^{y_0(1+x/x_+)}\frac{dy}{\sqrt{\delta ^2+y^2}}\nonumber\\
&~&+~\frac{1}{2}~\int _{x_+}^{\infty}\frac{dx}{\sqrt{\delta ^2+x^2}}\int _{-y_0(1-x/x_+)}^{y_0(1+x/x_+)}\frac{dy}{\sqrt{\delta ^2+y^2}}~.
\eea 
In the limiting case $\delta =0$, the only singularity of the integrand is at the
point $(x_+,0)$. It is an integrable one, hence we can perform the limit under
the integral. Then with the integrand $\frac{1}{xy}$ the integration becomes trivial
\beq
I^{\mbox{\scriptsize strip}}~=~\frac{\pi ^2}{4}+o(1)~.
\label{Istrip}
\eeq 
\\[10mm]
 
\end{document}